\documentclass{article}
\usepackage{PRIMEarxiv}

\usepackage[utf8]{inputenc} 
\usepackage[T1]{fontenc}    
\usepackage{newtxtext,newtxmath}
\usepackage{textcomp}
\usepackage[scaled=0.92]{helvet}          
\usepackage[scaled=0.84]{beramono}        
\usepackage{microtype}      

\usepackage{booktabs}       
\usepackage{amsfonts}       
\usepackage{nicefrac}       

\usepackage[hidelinks]{hyperref}       
\usepackage{xurl}           
\usepackage{lipsum}
\usepackage{fancyhdr}       
\usepackage{graphicx}
\graphicspath{{media/}}     

\usepackage{siunitx}
\usepackage{orcidlink}
\usepackage{setspace}
\title{Cross-lingual Comparison of Research Funding Projects with Multilingual Sentence-BERT : \\Evidence from KAKENHI, NIH, NSF, and UKRI
}

\author{
  Miki KIMURA-IDA 
  \orcidlink{0009-0005-4830-4657}
  \\
  National Institute of Science and Technology Policy (NISTEP), \\
  Ministry of Education, Culture, Sports, Science and Technology (MEXT) \\
  Tokyo, JAPAN\\
  \texttt{m-ida@nistep.go.jp}
}

\begin{document}
\maketitle

\begin{abstract}
Cross-national comparison of research funding projects is increasingly important for science policy and strategic planning, but language differences remain a major obstacle.
In particular, KAKENHI project descriptions are written primarily in Japanese, whereas projects from major overseas funding agencies, such as NSF, NIH, and UKRI, are documented in English.

This study investigates whether multilingual sentence embeddings can support meaningful cross-lingual comparison of research funding projects, with particular attention to the semantic effects of translating Japanese texts into English.
For each KAKENHI project, we construct two representations: the original Japanese text and its machine-translated English version, both embedded in a shared semantic space using a multilingual Sentence-BERT model.
We then compare their distances and nearest-neighbor relationships with respect to projects from English-language funding agencies.

The results show that the Japanese and translated English representations of the same KAKENHI project are, on average, located closer to one another than to native English projects, indicating substantial cross-lingual alignment.
However, the overlap of nearest neighbors between the two representations is limited, averaging \num{2.9} out of \num{10}.
This suggests that multilingual embeddings capture semantic similarity across languages to a meaningful extent, while language differences and translation still affect the local structure of the embedding space.

These findings suggest that multilingual embeddings provide a useful basis for large-scale exploratory comparison of funding projects across countries and agencies.
At the same time, they offer an empirical reference for assessing semantic drift when Japanese research project data are translated into English for international analysis.
\end{abstract}

\keywords{multilingual embeddings \and cross-lingual comparison \and research funding \and semantic drift}

\section{Introduction}
\label{sec:Introduction}
Comparative analysis of research funding systems plays a crucial role in understanding global research trends, identifying emerging scientific fields, and informing science and technology policy.
In recent years, increasing attention has been paid to cross-national comparisons of funding projects across major research systems.
However, a fundamental challenge arises from language differences in project descriptions.
In Japan, KAKENHI (Grants-in-Aid for Scientific Research), administered mainly by the Japan Society for the Promotion of Science (JSPS), is a major national competitive funding program for academic research.
Research projects funded under KAKENHI are primarily documented in Japanese, while projects funded by English-language funding agencies, such as the National Science Foundation (NSF), the National Institutes of Health (NIH) in the United States, and UK Research and Innovation (UKRI) in the United Kingdom, are described in English.
This linguistic discrepancy complicates the application of natural language processing (NLP) techniques for direct comparison.
A straightforward solution is to translate Japanese texts into English using machine translation.
While effective in some contexts, this approach introduces several issues.
Machine translation may alter the original meaning, particularly in specialized scientific domains, and may introduce systematic biases or artifacts that affect downstream analysis.
Recent advances in multilingual representation learning offer an alternative.
Models such as Sentence-BERT (sBERT) \cite{reimers-2019-sentence-bert} have been extended to multilingual settings, enabling texts in different languages to be embedded into a shared vector space.
In principle, semantically similar texts—regardless of language—should be located near each other in this space.
This study explores the feasibility and limitations of such an approach in the context of research funding data.
Specifically, we address the following research questions:

\begin{center}
\begin{minipage}{0.8\linewidth}
\begin{description}
\item [RQ1:] To what extent can multilingual embeddings enable cross-lingual comparison of research projects without machine translation?
\item [RQ2:] How consistent are local neighborhood structures across languages?
\end{description}
\end{minipage}
\end{center}

\section{Related Work}
\label{sec:Related Works}
Multilingual sentence embeddings have been widely studied in recent years, particularly in the contexts of cross-lingual information retrieval, semantic similarity, and machine translation evaluation.
Models based on multilingual transformers aim to align textual representations across languages through shared training objectives, making it possible to compare semantically related texts within a common vector space.
Sentence-BERT further advances this line of work by optimizing sentence embeddings for semantic similarity tasks, and multilingual variants of Sentence-BERT have shown strong performance in cross-lingual retrieval and matching tasks.

Research on cross-lingual retrieval has also demonstrated the practical importance of directly comparing Japanese and English texts.
For example, \cite{asai-etal-2021-xor} presents a multilingual retrieval setting that explicitly addresses cross-lingual information access between languages such as Japanese and English, highlighting the importance of robust representation learning for retrieval across language boundaries.
This line of work suggests that multilingual embedding models can provide a useful basis for comparing semantically related texts even when they are written in different languages.

Recent studies have further improved the quality of cross-lingual sentence representations and their applicability to multilingual comparison tasks.
In this context, \cite{hoshino-etal-2024-cross} provides an important recent contribution by demonstrating the effectiveness of cross-lingual semantic representation methods for multilingual text analysis.
Such studies support the view that multilingual sentence embeddings offer a promising foundation for analyzing semantic relationships across languages in a consistent and scalable manner.

In parallel, prior studies have explored the use of distributed representations for analyzing scientific texts, including research papers, abstracts, and project descriptions.
Earlier approaches often relied on word embeddings such as FastText\cite{joulin2016fasttext}
\footnote{\url{https://fasttext.cc/}}
and constructed document-level representations by aggregating word vectors, whereas more recent methods directly encode sentence- or document-level semantics using transformer-based architectures.
These developments have improved the ability to capture semantic similarity beyond surface-level lexical overlap.

However, relatively few studies have systematically examined cross-lingual comparisons of research funding data at scale, particularly in the context of science and innovation policy analysis.
While multilingual embedding methods have been actively studied in information retrieval and general semantic similarity tasks, their application to large-scale funding databases remains limited.
This study addresses that gap by conducting an empirical analysis of research project data from KAKENHI, NIH, NSF, and UKRI using multilingual sentence embeddings and nearest-neighbor analysis.

\section{Materials and Methods}
\label{sec:Method}
This study investigates the cross-lingual comparability of research project texts by representing each project description as a dense vector in a shared semantic space and then examining the neighborhood structure of those vectors.
The target corpora consist of research project records collected from KAKENHI, NIH, NSF, and UKRI for the period from about FY2017 to FY2025.
For each project, the textual information used for analysis consists of the project title and abstract.
Across all funding agencies, multiple funding categories, programs, and schemes are included together in a unified dataset and are not analyzed separately by scheme.
Thus, the analysis focuses on semantic relationships across the full set of projects rather than on differences among funding categories or program types.

The final dataset contains 
\num{264881} KAKENHI projects, 
\num{137102} NIH projects, 
\num{107250} NSF projects, and 
\num{83336} UKRI projects.

\begin{table}[htb]
 \caption{Number of projects included in the dataset}
 \label{tab:Num of PJ}
  \centering
  \begin{minipage}{0.3\linewidth}
      \begin{tabular}{lp{8mm}r}
        \toprule
        Funding     & & Counts\\
        \midrule
            \textbf{KAKENHI}   & & \num{264881} \\
            UKRI             & & \num{83336} \\
        \bottomrule
      \end{tabular}
  \end{minipage}
  \hspace{1mm}
  \begin{minipage}{0.3\linewidth}
      \begin{tabular}{lp{8mm}r}
        \toprule
        Funding     & & Counts\\
        \midrule
            NIH              & & \num{137102} \\
            NSF              & & \num{107250} \\
        \bottomrule
      \end{tabular}
  \end{minipage}
\end{table}

To evaluate the effect of language and translation, two coordinate types are prepared for KAKENHI: the original Japanese text and its machine-translated English version.
For the main analysis, the MT sample was restricted to projects for which both the title and the abstract had been translated by the LLM.
If the source text was already written in English, it was retained without modification in the author-written English baseline.
Translation of Japanese texts was performed  
primarily using AWS Bedrock with Claude Sonnet 4.5\footnote{\texttt{anthropic.claude-sonnet-4-5-20250929-v1}}.
In the broader preprocessing stage, a small number of texts that could not be processed by the LLM because of safety restrictions were translated using AWS Translate; however, such cases were excluded from the main MT analysis.
For embedding, the title and abstract of each project were concatenated into a single text string separated by a space.
For machine translation, the title and abstract were translated separately, with each field submitted independently using the same prompt.
Thus, if a project had both a title and an abstract, two translation calls were made for that project.
Missing values were handled by pairwise deletion, and the sampled projects used in the main analysis contained all fields required for the relevant comparisons.
No manual post-editing was applied.
For the overseas funding agencies, native English project texts are used as the comparison set.
All texts are converted into \num{384}-dimensional sentence embeddings using the multilingual Sentence-BERT model 
\texttt{paraphrase-multilingual-MiniLM-L12-v2}.
This model maps semantically similar texts to nearby positions in a high-dimensional vector space, thereby enabling cross-lingual comparison within a unified embedding space.
In this study, two complementary evaluation measures are introduced: distance and neighborhood overlap.

\begin{figure}
    \begin{minipage}{0.49\linewidth}
      \centering
      \includegraphics[width=0.9\linewidth]{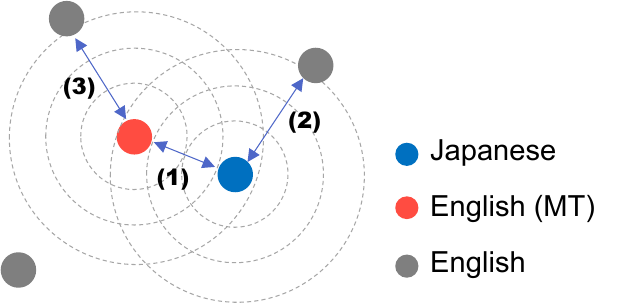}
      \caption{Distance metrics by coordinate type}
      \label{fig:Distance metrics by coordinate type}
    \end{minipage}
    \begin{minipage}{0.49\linewidth}
      \centering
      \includegraphics[width=0.9\linewidth]{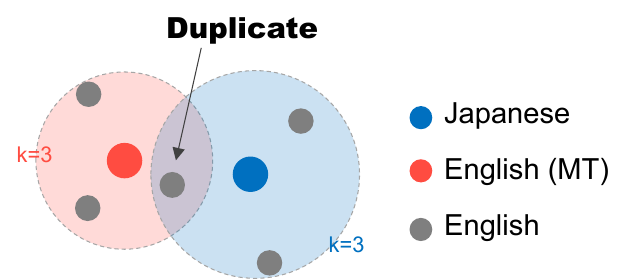}
      \vspace{3.2mm}
      \caption{k-NN overlap between coordinate types}
      \label{fig:k-NN overlap between coordinate types}
    \end{minipage}
\end{figure}

In distributed representations, semantic similarity is interpreted in terms of distance: semantically similar items are expected to be located near one another, whereas semantically dissimilar items are expected to be located farther apart.
On the basis of this assumption, the present study evaluates the embeddings using distance-based measures.
First, for each KAKENHI project, the distance between the original Japanese representation and the machine-translated English representation is calculated.
In order to determine whether this distance should be regarded as relatively small or large, distances to semantically related projects are also calculated as a baseline.
In the present setting, each KAKENHI project has two representations, namely the original-language version and the translated version.
For each of these two representations, the nearest native English projects are retrieved from NIH, NSF, and UKRI, and the distances to those neighboring projects are used as baseline references.
As illustrated in Figure~\ref{fig:Distance metrics by coordinate type}, three types of distances are examined: 
(1) the distance between the original-language and translated representations of the same KAKENHI project, 
(2) the distance between the original-language KAKENHI representation and native English projects, and 
(3) the distance between the translated KAKENHI representation and native English projects.
At the same time, even when the distance between the original-language and translated representations is not particularly small, the two representations may still be regarded as practically similar if they are surrounded by the same neighboring projects.
For this reason, in addition to distance, a second indicator based on neighborhood overlap is set.
More specifically, this overlap serves as a measure of the extent to which local semantic neighborhoods are preserved when the same project is represented in different languages.
A larger overlap indicates that the neighborhood structure is more stable across languages, whereas a smaller overlap suggests that translation or language differences alter the local topology of the embedding space.
In the present analysis, the top $k$ native English neighbors are retrieved for both the original-language and translated representations of the same KAKENHI project, and the degree of overlap between the two neighbor sets is measured.
A schematic example with $k=3$ is shown in Figure~\ref{fig:k-NN overlap between coordinate types}.

For visual inspection of cross-lingual alignment, the high-dimensional embeddings are projected onto two dimensions using UMAP \cite{mcinnes2020}.
This visualization is restricted to KAKENHI projects and displays, for each project, both the original Japanese representation and its machine-translated English representation.
The two-dimensional projection is used only for qualitative interpretation, whereas the quantitative evaluation is based on distances and neighborhood relations in the original embedding space.
When the cross-lingual alignment is high, the Japanese and English representations of the same project are expected to be mapped to nearly identical positions, causing points of different colors to overlap visually and appear almost as a single point.
Conversely, visible separation between the two points suggests a weaker alignment between the two language representations.

It should be noted that the retrieval of neighboring points is computationally simplified using the technique described as follows.
In principle, nearest-neighbor retrieval requires exhaustive distance computation between each query vector and all vectors in the dataset.
However, this procedure is essentially of quadratic complexity, that is, $O(N^2)$, and is computationally expensive for a corpus of this scale.
To address this problem, the present study employs NGT (Neighborhood Graph and Tree), an approximate nearest-neighbor search method developed by \cite{iwasaki2018}
\footnote{\url{https://github.com/NGT-labs/NGT}}.
 NGT accelerates retrieval by indexing the vector space using graph and tree structures, thereby enabling scalable neighborhood analysis for large-scale data.
 Although the retrieved neighbors are approximate rather than exact, this approach provides a practical and scalable solution for semantic neighborhood analysis in a large collection of multilingual project embeddings.

\section{Analysis}
\label{sec:Analysis}
The analysis examines how research projects are positioned across coordinate types in the shared embedding space and what these positional relationships imply for cross-national comparison of funding projects.
In particular, we focus on whether the original Japanese descriptions of KAKENHI projects, their machine-translated English counterparts, and native English project descriptions from overseas funding agencies form similar local and global structures in the embedding space.

For the main analysis, the sample was restricted to KAKENHI projects for which both the title and the abstract had been translated by the LLM.
From this subset, \num{1000} project IDs were randomly sampled, and the top \num{10} nearest-neighbor projects from NSF, NIH, and UKRI were retrieved for each sampled project using NGT.
As a baseline for comparison, an additional \num{1000} KAKENHI projects for which both the title and the abstract were originally written in Japanese and English were sampled separately and analyzed in the same way.
The following analyses are based on these randomly selected IDs.
 
First, the distance-based results indicate that the average distance between the original Japanese text and the machine-translated English text of the same KAKENHI project is smaller than the average distance between KAKENHI-derived texts and native English project texts.
In this analysis, distances are measured as Euclidean (L2) distances in the original \num{384}-dimensional embedding space generated by the multilingual Sentence-BERT model
\footnote{Each vector is normalized to unit length.}.
As shown in Table~\ref{tab:distance llm}, the average distance between the native Japanese and machine-translated English representations of the same project is \num{0.62}, whereas the corresponding average distances between native Japanese and native English projects, and between machine-translated English and native English projects, are \num{0.75} and \num{0.76}, respectively.
This result suggests that the multilingual embedding model places the two language representations of the same project relatively close to one another, while still preserving a non-negligible difference between them.
At the same time, the fact that the Japanese and machine-translated English versions do not collapse into identical positions indicates that language and translation continue to affect semantic positioning in the embedding space.
In other words, multilingual sentence embeddings support cross-lingual comparability, but they do not completely eliminate language-related variation.
This point is important in the context of funding portfolio analysis, because even small positional differences may alter how projects are situated relative to surrounding thematic clusters.

Second, the average neighborhood overlap was \num{2.2} for the baseline and \num{2.9} for the machine-translated version, out of the top \num{10} neighbors (Table~\ref{tab:duplication}).
Therefore, the overlap is limited, indicating that the local neighborhood structure is not fully preserved when the same KAKENHI project is represented in different languages.
This means that even when two representations of the same project are relatively close in terms of average distance, they may still be surrounded by somewhat different sets of neighboring projects.
For funding portfolio analysis, this result implies that comparisons based on local thematic context remain sensitive to language choice and translation.
This observation is consistent with the interpretation that multilingual embeddings are useful for exploratory cross-lingual comparison at the portfolio level, but that local semantic relations should be interpreted with caution.
If the neighboring projects differ depending on whether Japanese or translated English text is used, then the apparent thematic affiliation of a project may also shift slightly.
Such shifts do not invalidate cross-lingual comparison, but they do indicate that the local topology of the embedding space is influenced by both the multilingual model and the translation process.

\begin{figure}
  \centering
  \fbox{\includegraphics[width=0.68\linewidth]{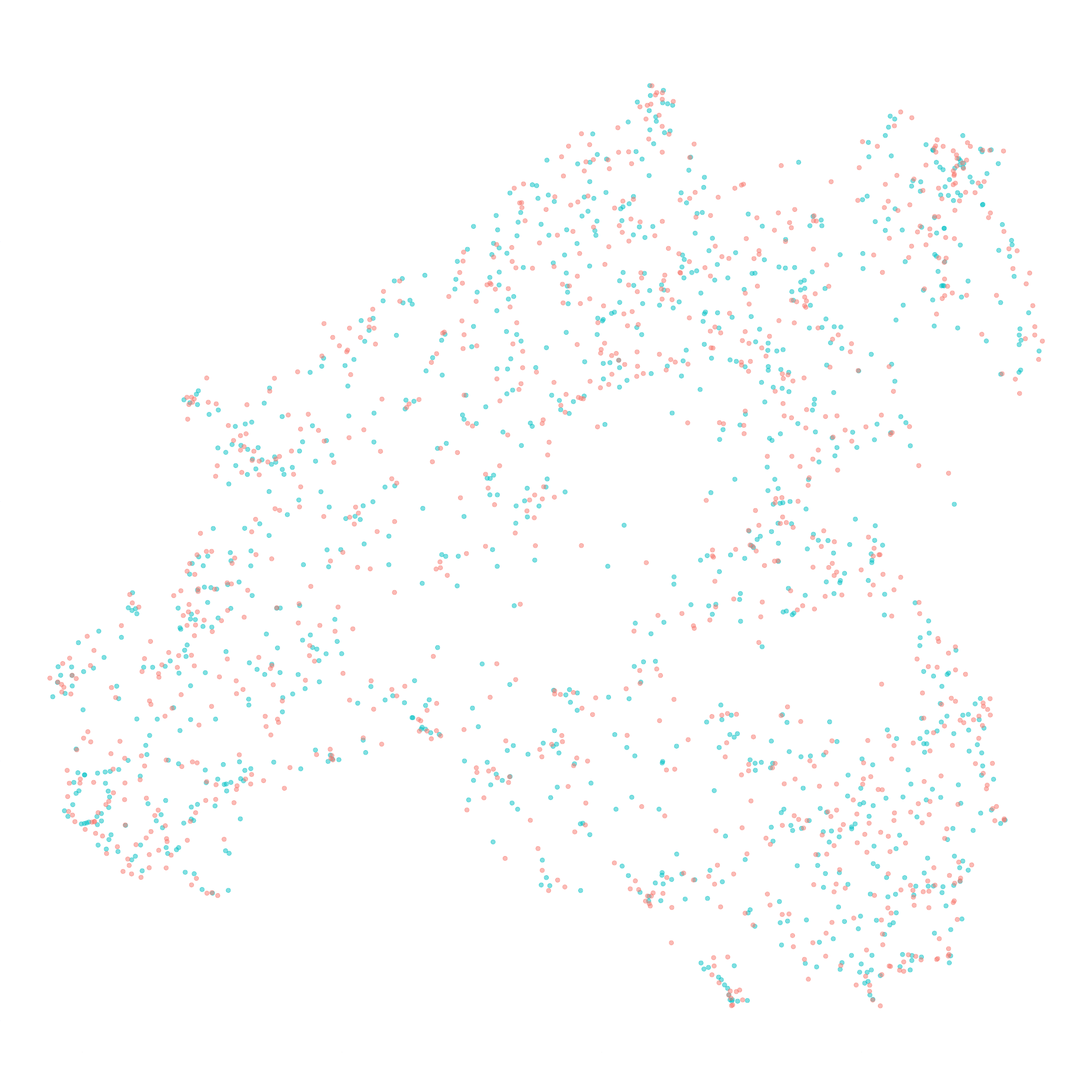}}
  \caption{UMAP projection of KAKENHI project embeddings in Japanese (blue) and machine-translated English (red)}
  \label{fig:umap_kaken_en_jp}
\end{figure}

Third, the UMAP visualisation in Figure~\ref{fig:umap_kaken_en_jp} provides a qualitative view of the global structure of the embedding space.
Although UMAP does not preserve all pairwise distances exactly, it is useful for observing broad distributional tendencies.
If the Japanese and English representations are largely intermixed, this suggests that the multilingual model succeeds in mapping them into a shared semantic space.
At the same time, incomplete overlap or visible separation between point clouds would indicate that residual language-dependent structure remains even after multilingual embedding.
Such a pattern would be consistent with the distance and neighbor-overlap results described above.

Taken together, these findings address the first two research questions in a complementary manner.
Regarding RQ1, the results suggest that multilingual sentence embeddings provide a practical basis for cross-lingual comparison of research funding projects without fully relying on translation, although language-related variation remains.
The shared embedding space appears sufficiently aligned to support broad comparison of funding projects across KAKENHI, NIH, NSF, and UKRI.
Regarding RQ2, however, the findings also indicate that local neighborhood structures are not fully consistent across language representations.
Both language differences and translation effects continue to influence project positioning and neighborhood relationships at the local level.
Accordingly, multilingual embeddings are well suited to exploratory and structural analyses of funding projects, including comparisons of thematic breadth, concentration, and relative positioning across funding agencies, but they should be interpreted carefully in analyses that require fine-grained local semantic precision.

\begin{table}[bhtp]
 \caption{Average Euclidean distances across coordinate types (Machine Translate)}
 \label{tab:distance llm}
  \centering
  \begin{tabular}{lrrr}
    \toprule
    Item     & Distance (Avg.) & SD ($s$) & Var ($s^2$)\\
    \midrule
        Between Native Japanese and MT English (Same Project)
        & \num{0.62} & \num{0.12} & \num{0.02} \\
        Between Native Japanese and Native English
        & \num{0.76} & \num{0.08} & \num{0.01} \\
        Between MT English and Native English
        & \num{0.75} & \num{0.10} & \num{0.01} \\
    \bottomrule
    \multicolumn{4}{r}{\small (n=\num{1000}, random sampling)}
  \end{tabular}
\end{table}

\begin{table}[bhtp]
 \caption{Average Euclidean distances across coordinate types (Baseline)}
 \label{tab:distance baseline human}
  \centering
  \begin{tabular}{lrrr}
    \toprule
    Item     & Distance (Avg.) & SD ($s$) & Var ($s^2$)\\
    \midrule
        Between Native Japanese and English* (Same Project)
        & \num{0.71} & \num{0.16} & \num{0.03} \\
        Between Native Japanese and Native English
        & \num{0.76} & \num{0.08} & \num{0.01} \\
        Between English* and Native English
        & \num{0.76} & \num{0.10} & \num{0.01} \\
    \bottomrule
    \multicolumn{4}{r}{\small * KAKENHI with author-written English; (n=\num{1000}, random sampling)}
  \end{tabular}
\end{table}

\begin{table}[bhtp]
 \caption{Average neighborhood overlap across coordinate types}
 \label{tab:duplication}
  \centering
  \begin{tabular}{lr}
    \toprule
    Item     & Average overlap ($k=10$)\\
    \midrule
        Between Native Japanese and Machine Translated English & \num{2.9} \\
        Between Native Japanese and Author-written English & \num{2.2} \\
    \bottomrule
    \multicolumn{2}{r}{\small (n=\num{1000} for each group, random sampling)}
  \end{tabular}
\end{table}

\begin{figure}[bhtp]
  \centering
  \includegraphics[width=0.8\linewidth]{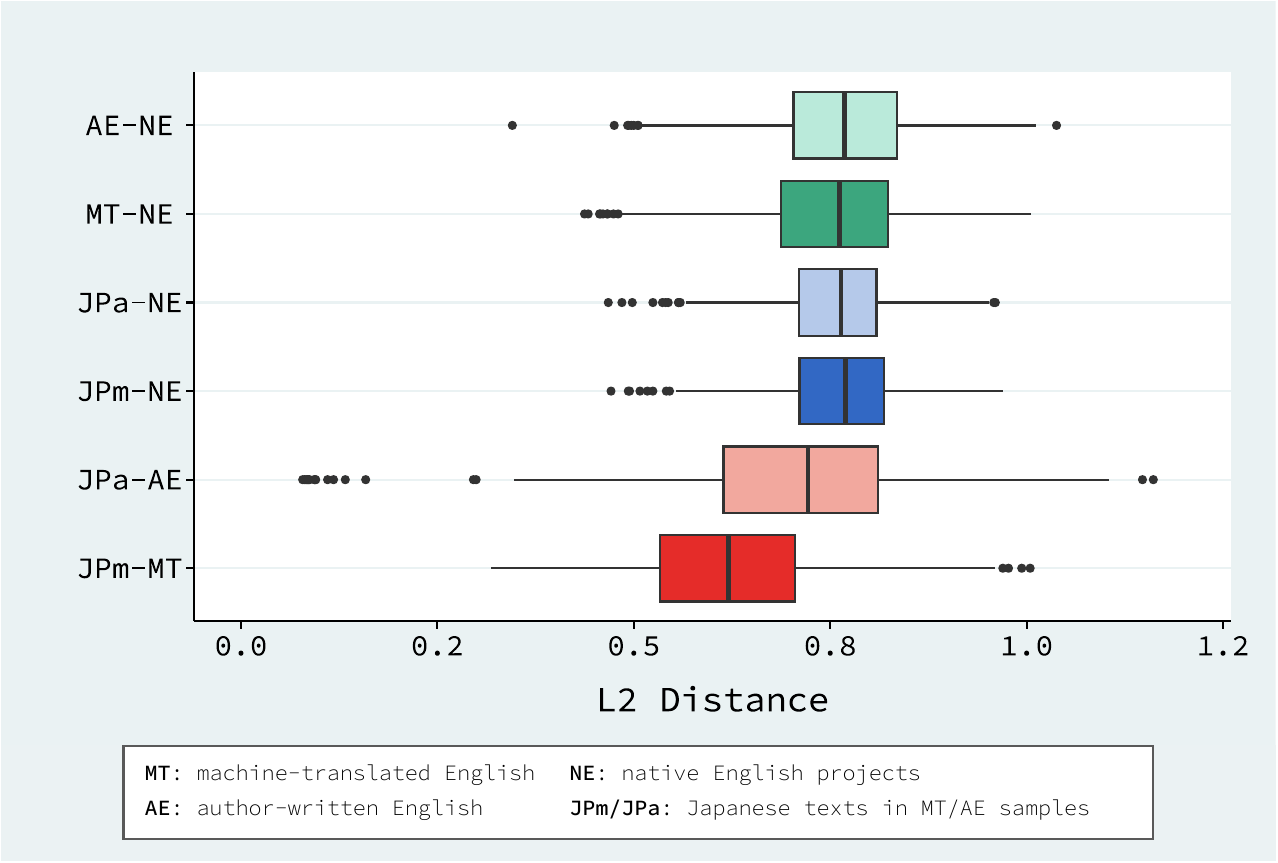}
  \caption{Distribution of normalized L2 distances}
  \label{fig:distribution}
\end{figure}

\section{Discussion}
\label{sec:Discussion}
The present study examined whether multilingual sentence embeddings can support cross-lingual comparison of research funding projects without relying entirely on machine translation.
The results suggest that such embeddings provide a meaningful shared semantic space, although this space is not fully language-invariant.
In particular, the original Japanese and English representations of the same KAKENHI project are, on average, located closer to each other than either is to native English projects, indicating substantial cross-lingual alignment.
At the same time, the relatively limited overlap in local neighborhoods suggests that this alignment remains imperfect, especially at the level of fine-grained local structure.
An additional analysis shows that the distances from KAKENHI projects to nearby English-language projects are broadly similar whether the English text is author-written or generated by LLM-based translation.
By contrast, the distance between the Japanese and English representations of the same project is smaller for the LLM-translated group than for the author-written English group, and the neighborhood overlap is also slightly higher in the LLM-translated group.

This suggests that, within the embedding space used in the present study, LLM-based translation may preserve the semantic position of the original Japanese text more faithfully than author-written English descriptions.
However, this should not be interpreted as a general claim about translation quality, but rather as a result specific to alignment within the present representation space.

At the same time, author-written English is not available for all projects, and the composition of the English-side data therefore remains constrained.
For broad, structural comparisons across funding agencies, multilingual embeddings appear sufficiently robust to support analysis of overall thematic distributions and relative positioning among agencies.
This is consistent with the broader aim of the study, which is to place projects from KAKENHI, NIH, NSF, and UKRI in a common semantic space and compare their distributional structures without depending on existing field classifications.

However, the results also indicate a clear limitation.
If the nearest-neighbor overlap between two representations of the same project is only \num{2.9} out of \num{10} on average, then local thematic interpretation remains sensitive to language choice and translation.
Although a project may be placed in roughly the same region of the embedding space across languages, its immediate semantic context can still shift.
Accordingly, the method is better suited to aggregate and meso-level analysis than to exact local matching.

Accordingly, part of the observed differences in cross-lingual distances and neighborhood structures may reflect properties of the translation systems themselves, rather than only the behavior of the multilingual embedding model.
This issue affects not only reproducibility but also the generalizability and interpretability of the results.
The comparison between original Japanese and translated English suggests that translation is not simply a neutral preprocessing step.
It can move project representations in the embedding space and thereby alter their local relationships to surrounding projects.
At the same time, multilingual embeddings and translation should be seen not as mutually exclusive alternatives, but as different ways of balancing scalability and semantic precision.
Future work should therefore examine this issue more closely by aligning the target years more strictly and conducting subset analyses that explicitly control for temporal coverage.
The central contribution of this study lies in showing that multilingual sentence embeddings can support cross-lingual comparison of research project texts.
A secondary implication is that, once projects from different countries are placed in a shared semantic space, their thematic breadth, concentration, and relative positioning can be compared without relying entirely on agency-specific classification systems.
This provides a useful perspective for examining how KAKENHI is positioned relative to overseas funding portfolios with different institutional orientations, including more mission-oriented agencies such as NIH and UKRI and broader basic-science support structures such as NSF.

\section{Conclusion}
\label{sec:Conclusion}
This study examined the extent to which multilingual sentence embeddings can support cross-lingual comparison of research funding projects, with particular attention to the semantic effects of translating Japanese project texts into English.
Using project data from KAKENHI, NIH, NSF, and UKRI, and comparing the original Japanese and machine-translated English representations of KAKENHI projects, the analysis showed that multilingual embeddings capture cross-lingual semantic similarity to a meaningful extent.
The original Japanese and translated English versions of the same project were, on average, located closer to one another than to native English projects, indicating substantial cross-lingual alignment.

At the same time, the overlap of local neighborhoods was limited, suggesting that language and translation continue to affect the fine structure of the embedding space.
This implies that multilingual embeddings are sufficiently robust for large-scale comparative analysis, but not fully invariant to language choice when more fine-grained semantic relationships are concerned.

The contribution of this study can therefore be understood from two complementary perspectives.
First, it provides a methodological framework for placing project texts from different funding agencies into a shared semantic space without relying on predefined classification systems.
This makes it possible to compare funding projects across countries and agencies in a flexible and scalable manner.
Second, and more importantly, the study provides an empirical reference point for assessing the degree of semantic drift introduced by translating Japanese research project texts into English.
In this sense, the analysis is not only about whether multilingual comparison is feasible, but also about how much meaning may change when Japanese funding data are Englishized for international comparison.

This point is particularly important because language barriers often become more explicit in downstream analytical tasks.
While multilingual embeddings may allow project texts in different languages to be compared within a shared semantic space, other approaches, such as topic extraction using LDA, are generally much more sensitive to lexical and linguistic differences.
From this perspective, the present results offer a useful benchmark for understanding how far English translation can preserve the semantic position of Japanese project texts before applying downstream analyses in an English-based setting.

Overall, the findings suggest that multilingual embeddings provide a promising basis for cross-national and policy-oriented analysis of research funding projects, especially when the objective is exploratory or structural comparison at scale.
At the same time, the results indicate that machine translation is not semantically neutral, and that its effects should be taken into account when Japanese project data are converted into English for subsequent analysis.
The approach presented here is therefore best understood not as a tool for exact semantic matching, but as a practical framework for comparative analysis and as a reference for evaluating semantic change associated with translation in multilingual research funding data.

\section*{Acknowledgments}

The author would like to express sincere gratitude to Dr. Koshiba, Senior Research Fellow at National Institute of Science and Technology Policy, for his generous support, insightful advice, and valuable comments throughout the course of this paper.
The author is also grateful to all those who provided helpful suggestions on the research framework, data preparation, and interpretation of the findings.

Any remaining errors or omissions are solely the responsibility of the author.

\addcontentsline{toc}{section}{References}
\newcommand{\etalchar}[1]{$^{#1}$}

\end{document}